\documentclass[%
 reprint,
superscriptaddress,
 amsmath,amssymb,
 aps,
]{revtex4-2}

\usepackage{graphicx}
\usepackage{dcolumn}
\usepackage{bm}

\usepackage{graphicx}
\usepackage{amsmath}
\usepackage{amssymb}
\usepackage[usenames, dvipsnames]{xcolor}
\usepackage[colorlinks=true,citecolor=black,linkcolor=black,urlcolor=black]{hyperref}

\usepackage{tabularx}
\usepackage{color}
\usepackage{xspace}
\usepackage{hyperref}

\begin{document}

	\title{Superconducting density of states and vortex lattice of LaRu$_2$P$_2$ observed by Scanning Tunneling Spectroscopy}%

	\author{Marta Fern\'andez-Lomana}
    \affiliation{Laboratorio de Bajas Temperaturas y Altos Campos Magnéticos, Departamento de Física de la Materia Condensada, Instituto Nicolás Cabrera and Unidad Asociada UAM-CSIC, Universidad Autónoma de Madrid, E-28049, Madrid, Spain}
    \affiliation{Condensed Matter Physics Center (IFIMAC),  Universidad Autónoma de Madrid, E-28049, Madrid, Spain}
	
	\author{Paula Obladen Aguilera}
 \affiliation{Laboratorio de Bajas Temperaturas y Altos Campos Magnéticos, Departamento de Física de la Materia Condensada, Instituto Nicolás Cabrera and Unidad Asociada UAM-CSIC, Universidad Autónoma de Madrid, E-28049, Madrid, Spain}
    \affiliation{Condensed Matter Physics Center (IFIMAC),  Universidad Autónoma de Madrid, E-28049, Madrid, Spain}

		\author{Beilun Wu}
 \affiliation{Laboratorio de Bajas Temperaturas y Altos Campos Magnéticos, Departamento de Física de la Materia Condensada, Instituto Nicolás Cabrera and Unidad Asociada UAM-CSIC, Universidad Autónoma de Madrid, E-28049, Madrid, Spain}
    \affiliation{Condensed Matter Physics Center (IFIMAC),  Universidad Autónoma de Madrid, E-28049, Madrid, Spain}
	
		\author{Edwin Herrera}
	 \affiliation{Laboratorio de Bajas Temperaturas y Altos Campos Magnéticos, Departamento de Física de la Materia Condensada, Instituto Nicolás Cabrera and Unidad Asociada UAM-CSIC, Universidad Autónoma de Madrid, E-28049, Madrid, Spain}
    \affiliation{Condensed Matter Physics Center (IFIMAC),  Universidad Autónoma de Madrid, E-28049, Madrid, Spain}

\author{Hermann Suderow}
 \affiliation{Laboratorio de Bajas Temperaturas y Altos Campos Magnéticos, Departamento de Física de la Materia Condensada, Instituto Nicolás Cabrera and Unidad Asociada UAM-CSIC, Universidad Autónoma de Madrid, E-28049, Madrid, Spain}
    \affiliation{Condensed Matter Physics Center (IFIMAC),  Universidad Autónoma de Madrid, E-28049, Madrid, Spain}		
	
\author{Isabel Guillamón}
 \affiliation{Laboratorio de Bajas Temperaturas y Altos Campos Magnéticos, Departamento de Física de la Materia Condensada, Instituto Nicolás Cabrera and Unidad Asociada UAM-CSIC, Universidad Autónoma de Madrid, E-28049, Madrid, Spain}
    \affiliation{Condensed Matter Physics Center (IFIMAC),  Universidad Autónoma de Madrid, E-28049, Madrid, Spain}

	\date{\today}
	
	\begin{abstract}
We provide the superconducting density of states of the iron based superconductor LaRu$_2$P$_2$ (T$_c$= 4.1 K), measured using millikelvin Scanning Tunneling Microscopy. From the tunneling conductance, we extract a density of states which shows the opening of a s-wave single superconducting gap. The temperature dependence of the gap also follows BCS theory. Under magnetic fields, vortices present Caroli de Gennes Matricon states, although these are strongly broadened by defect scattering. From the vortex core size we obtain a superconducting coherence length of $\xi=50$ nm, compatible with the value extracted from macroscopic H$_{c2}$ measurements. We discuss the comparison between s-wave LaRu$_2$P$_2$ and pnictide unconventional multiple gap and strongly correlated Fe based superconductors.
\end{abstract}

\maketitle 

LaRu$_2$P$_2$ (T$_c$= 4.1 K) was the pnictide superconductor with the highest T$_c$ known before the discovery of high T$_c$ superconductivity in non-stoichiometric pnictide superconductors \cite{Jeitschko1987superconducting,Kamihara2008iron,Razzoli2012bulk}.

High T$_c$ superconductivity in the pinctides and iron based superconductors indeed often appears in doped compounds. A relevant and well examined example is the BaFe$_2$As$_2$ system, which is an antiferromagnet that becomes superconducting by doping with holes and with electrons \cite{Physrevlett.101.107006, Physrevlett.101.117004}. Isoelectronic compounds in its stoichiometric composition, such as BaRu$_2$As$_2$ or BaFe$_2$P$_2$, do not exhibit superconductivity\,\cite{Physrevb.81.174512,Jiang2009}.

LaRu$_2$P$_2$ is isoeletronic to LaFe$_2$As$_2$. Both crystallize in the tetragonal ThCr$_2$Si$_2$ structure and have three dimensional Fermi surfaces. However, whereas LaRu$_2$P$_2$ is a superconductor with a T$_c$ of 4.1 K, LaFe$_2$As$_2$ is not\,\cite{Moll2011quantum}. A similar case is found in the isoelectronic oxyde compounds LaRuPO and LaFePO. These materials show similar Fermi surfaces with strong 2D character. LaRuPO is non-superconducting whereas LaFePO is superconducting with a T$_c$ of 6 K\,\cite{Hamlin_2008}. The substitution of Fe-3d orbitals with the much less localized Ru-4d orbitals leads to an increased bandwidth, resulting in reduced electronic correlations and lower values of the effective mass\,\cite{Brouet2010significant,Doi:10.1143/jpsj.78.053705}.  However, in the compounds with 3D electronic properties, LaRu$_2$P$_2$ and LaFe$_2$As$_2$, superconductivity is found in the material with smaller electronic correlations suggesting that the origin of superconductivity in LaRu$_2$P$_2$ is different than in other pnictide superconductors\,\cite{Moll2011quantum}.

LaRu$_2$P$_2$ presents under pressure a transition to a collapsed tetragonal phase in which the c-axis lattice parameter decreases by as much as 10$\%$ \cite{Drachuck2017collapsed}. The collapsed tetragonal phase arises at low pressures in systems with large steric effects, such as CaFe$_2$As$_2$\,\cite{Physrevlett.101.057006}. The steric effects are partially lifted in CaKFe$_4$As$_4$ and in KFe$_2$As$_2$ compounds, which also show a transition to the collapsed tetragonal phase under pressure. Superconductivity remains with a higher T$_c$ in the collapsed tetragonal phase of KFe$_2$As$_2$\,\cite{Physrevb.91.060508} and is suppressed in the collapsed tetragonal phases of CaKFe$_4$As$_4$\,\cite{Physrevb.96.140501} and CaFe$_2$As$_2$\,\cite{Physrevb.93.245139}. LaRu$_2$P$_2$ under external pressure presents a collapsed tetragonal phase above 2 GPa which is not superconducting\,\cite{Foroozani2014hydrostatic, Li2016pressure}.

Band structure calculations, Angular Resolved Photoemission (ARPES), and de Haas van Alphen measurements all show that the electronic band structure of LaRu$_2$P$_2$ is very different than the one found in many other pnictide superconductors\,\cite{Razzoli2012bulk, Moll2011quantum}. The Fermi surface has a strong three-dimensional character and consists of two sets of sheets derived from the Ru-4d orbitals that form a donut with a small hole centered around the Z point, a warped electron cylinder at the corners of the Brillouin zone, and an additional open three-dimensional sheet derived from a combination of La and Ru-4d orbitals\,\cite{Razzoli2012bulk,Moll2011quantum}. The tubes at the corners of the Brillouin zone are the only zones of the Fermi surface that can be considered close to two-dimensional; the rest of the Fermi surface is largely three-dimensional. Mass renormalization is practically absent over the whole band structure, suggesting that superconductivity in LaRu$_2$P$_2$ is not related to electronic correlations but rather to electron-phonon coupling\,\cite{Jeitschko1987superconducting, Ying2010isotropic}.

\begin{figure}[h!]
		\centering
		\begin{center}
			\includegraphics[width = 1\columnwidth]{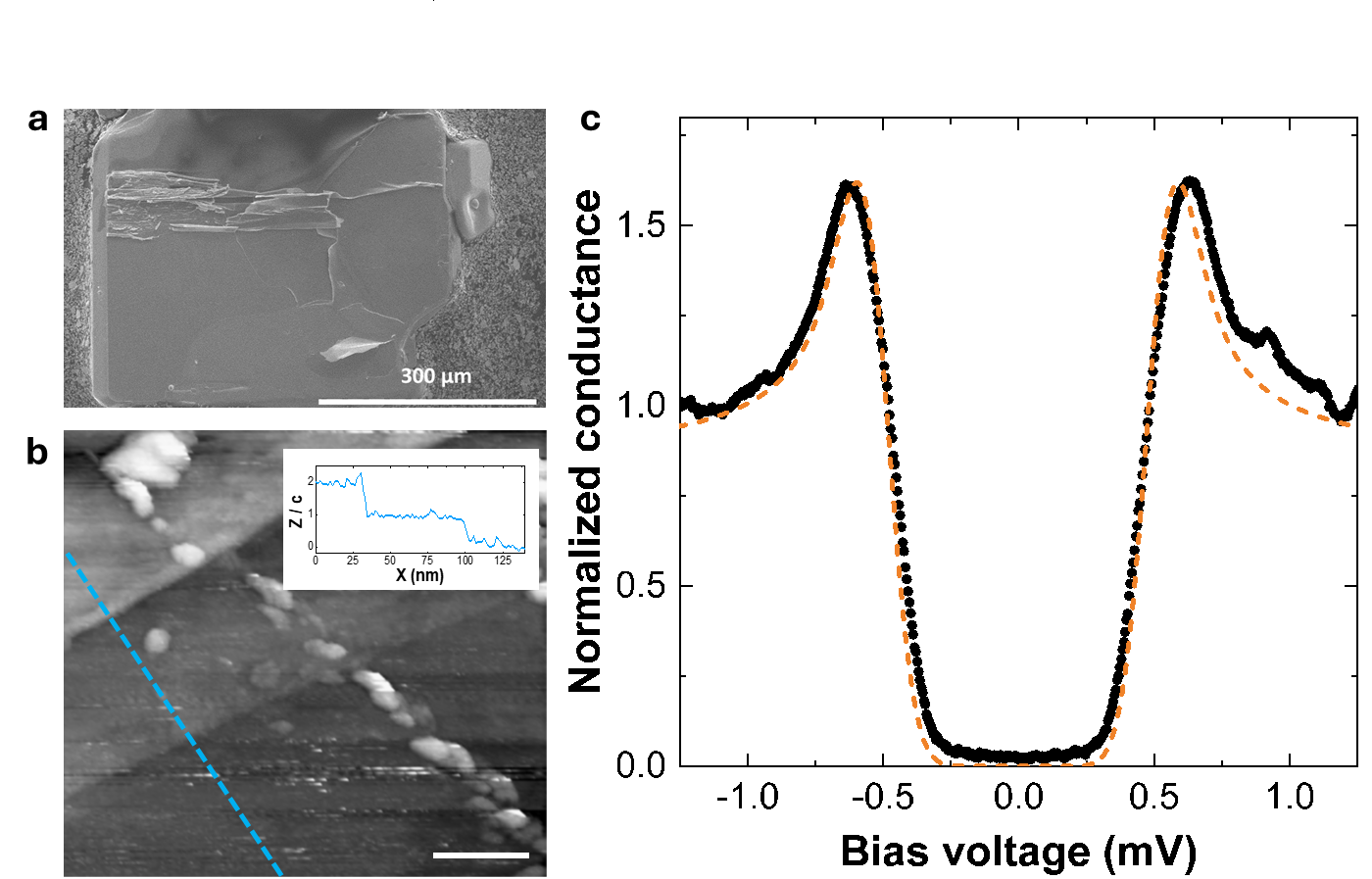}
		\end{center}
		\caption{\footnotesize{\textbf{a)} Scanning Electron Microscope (SEM) image of a LaRu$_2$P$_2$ sample after cryogenic cleaving and after removing the sample from the STM. \textbf{b)} STM image of the topography of the sample, taken with a bias voltage V=2.5 mV, a constant tunneling current I$_t$=2 nA and at T= 80 mK. In the inset we show the profile normalized to the c-axis lattice parameter of LaRu$_2$P$_2$, taken along the blue line. White scale bar is 40 nm long. \textbf{c)} Normalized conductance as a function of bias voltage measured at 80 mK (black) and a fit to the BCS theory (dashed line), with $\Delta$ = 0.61 meV.}\label{gapYtopo}}
\end{figure}

To better understand the role played by electron-phonon interactions it is fundamental to know the superconducting gap structure of LaRu$_2$P$_2$. Here we provide microscopic measurements of the superconducting gap of LaRu$_2$P$_2$ using Scanning Tunneling Microscopy (STM). We find a highly isotropic gap and study the vortex phase. We find that the superconducting coherence length is much larger than the coherence length of other iron pnictide superconductors. Our data establish LaRu$_2$P$_2$ as a nearly isotropic s-wave superconductor, contrasting anisotropic correlated electron superconductivity found in other pnictide superconducting materials.


  \begin{figure*}[ht]
		\centering
		\begin{center}
			\includegraphics[width = 0.8\textwidth]{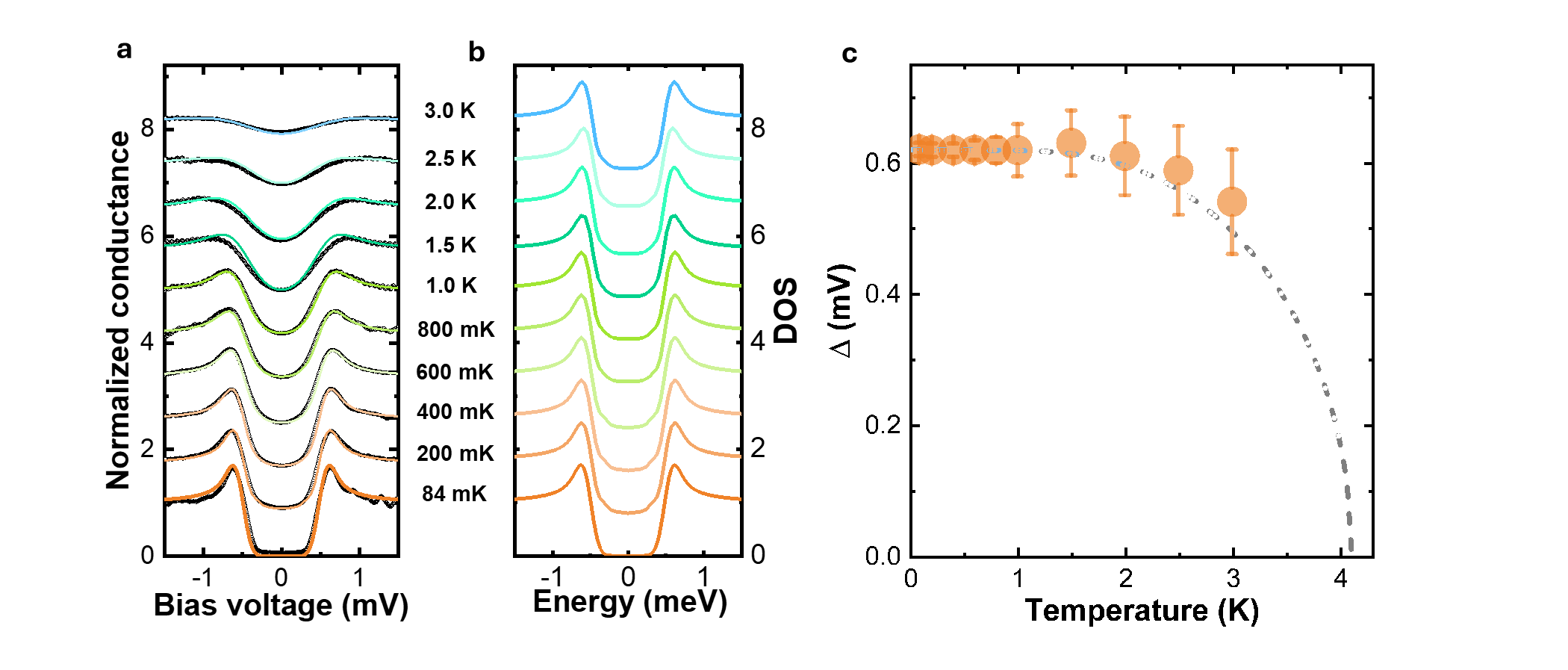}
		\end{center}
		\caption{\footnotesize{\textbf{a)} The normalized conductance as a function of bias voltage measured from 84\,mK to 3\,K is shown as black data points. The solid colored lines represent the convoluted density of states (DOS) with their respective temperatures. \textbf{b)} The DOS obtained from the curves in a) is plotted as a function of energy from 84\,mK to\,3 K. \textbf{c)} The temperature dependence of the superconducting gap is determined by the quasiparticle peak position of the DOS.}\label{DOSLa}}.
\end{figure*}

\begin{figure*}[ht]
		\centering
		\begin{center}
			\includegraphics[width = 0.73\textwidth]{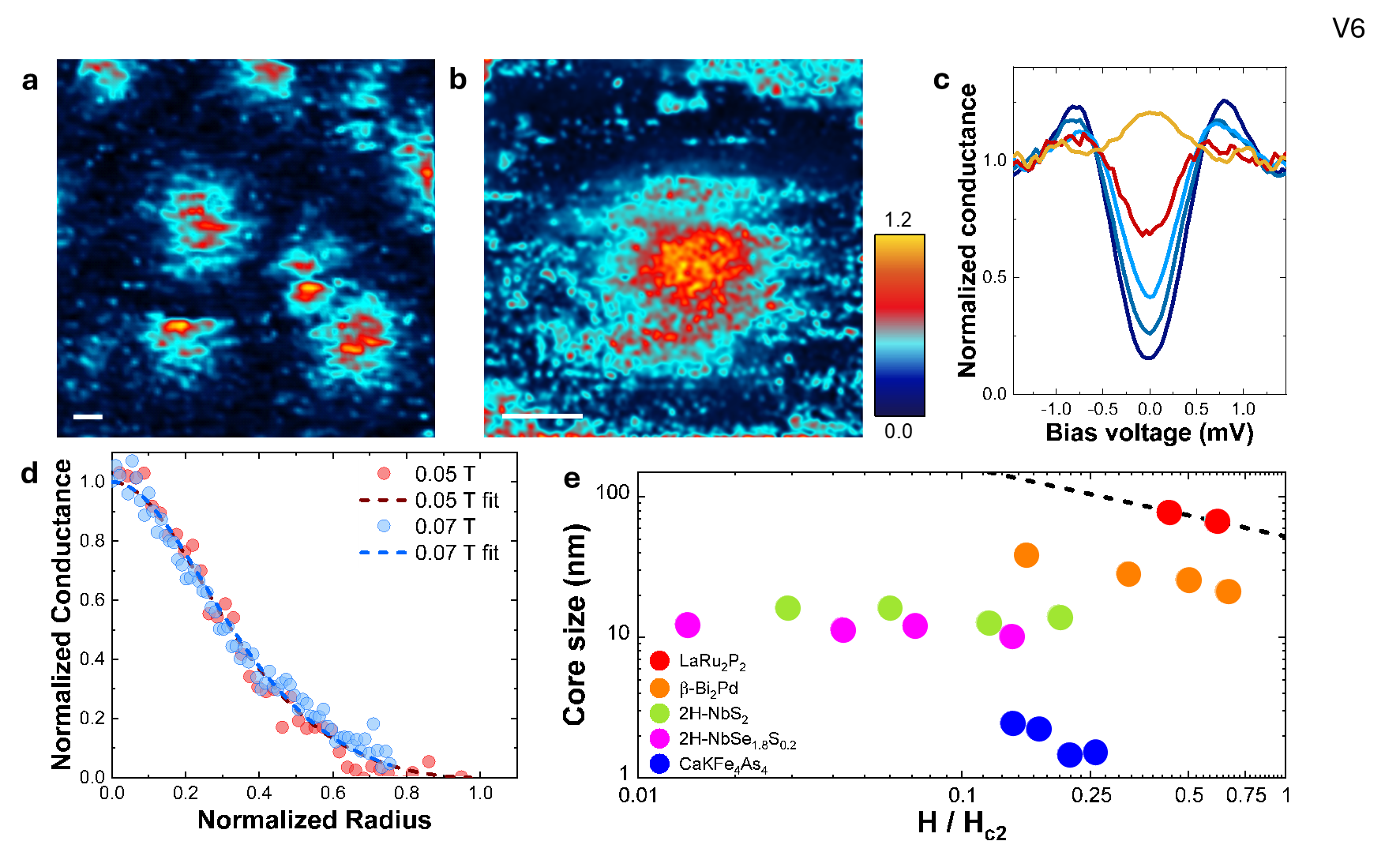}
		\end{center}		\caption{\footnotesize{\textbf{a)},\textbf{b)}  Zero bias tunneling conductance maps made at  magnetic fields 0.05\,T and 0.07\,T respectively. Images are taken at different locations at 90\,mK. White scale bar: 50 nm. The bar at the left provides the color scale in units of the normalized conductance. Vortex positions are  Delaunay triangulated. \textbf{c)} Normalized conductance along a line from inside to outside of a vortex. The conductance curves follow the same color code as the conductance map in a) and b). \textbf{d)} Normalized conductance vs distance from vortex centre normalized as\,\cite{Physrevb.97.134501}, at different magnetic fields. \textbf{e)} Vortex core size vs the magnetic field in LaRu$_2$P$_2$ compared to the core size in other superconductors ($\beta$-Bi$_2$Pd, 2H-NbS$_2$, 2H-NbSe$_{1.8}$S$_{0.2}$\,\cite{Physrevb.97.134501} and CaKFe$_4$As$_4$\,\cite{Physrevb.97.134501}). Dashed line shows the fit $C = \eta\sqrt{\Phi_o/\pi H}$, described in the text \cite{Physrevb.94.014517}}.\label{vortexLa}}
\end{figure*}

To perform the STM study of LaRu$_2$P$_2$, we cleaved up to five single crystalline samples at 4 K and under cryogenic vacuum conditions (see a picture of a sample in Fig.\,\ref{gapYtopo}a). Samples were grown using flux growth\,\cite{Canfield01}, as described in Ref.\,\cite{FernandezLomana2021}. It has been very difficult to obtain flat surfaces in LaRu$_2$P$_2$, in spite of searching more than thirty different fields of view using the in-situ positioning mechanism described in Ref.\,\cite{Suderow2011}. Being a three-dimensional crystal, the cleaving procedure might be complex and strain-free homogeneous surfaces might be difficult to create. Here we focus on the few particular surfaces where we found flat terraces and neat superconducting features. In Fig.\,\ref{gapYtopo}b, steps with a height of one unit cell can be seen. We use a STM described in Ref.\,\cite{uwu2} and the software described in Refs.\,\cite{Winspm,uwu1}. The resolution in energy is below 8 $\mu$eV and the lowest temperatures of our set-up is below 100 mK, as shown in Ref.\,\cite{uwu2}.

Fig.\,\ref{gapYtopo}c presents the measured conductance on one of these surfaces at 80 mK and 0 T. The conductance shows the opening of a superconducting gap with zero density of states at zero bias and well developed quasi-particle peaks. We have fitted our data using BCS expression (orange dashed line in Fig.\,\ref{gapYtopo}c) and obtain $\Delta=0.61$ meV which is $\Delta=1.72k_BT_c$ for T$_c$ = 4.1 K. This is in agreement with the T$_c$ obtained by macroscopic measurements \cite{Razzoli2012bulk,Moll2011quantum,Jeitschko1987superconducting, Ying2010isotropic}.

In Fig.\,\ref{DOSLa}a we show the normalized conductance as a function of bias voltage for different temperatures ranging from 84\,mK to 3\,K. Solid colored lines provide the convolution of the density of states with the derivative of the Fermi function. In Fig.\,\ref{DOSLa}b we show the density of states used for the convolution at each temperature, as described in Refs.\,\cite{SUDEROW2024600,Guillamon2008superconducting,uwu2,Physrevb.97.134501,PhysRevMaterials.7.024804,PhysRevResearch.4.023241,PhysRevB.92.054507}. The size of the energy gap is given by the position of the peaks in the density of states. The gap as a function of temperature is shown in Fig.\,\ref{DOSLa}c. The dashed line is the temperature dependence of the gap within BCS theory. We can see that the temperature dependence of the gap size extracted from the experiment follows well BCS theory.


Fig.\,\ref{vortexLa}a and b show conductance maps at 0\,mV for magnetic fields of 0.05\,T and 0.07\,T (approximately 0.43 H$_{c2}$ and 0.61 H$_{c2}$, with H$_{c2}$ = 0.127 T\,\cite{FernandezLomana2021}). In the 0.05\,T map we observe a few vortices. We see that vortices do not form an ordered triangular lattice, possibly due to the large intervortex separation and due to pinning. Nevertheless, the average intervortex distance is  213$\pm$50\,nm, which is compatible with the distance expected from Abrikosov theory for a triangular vortex lattice at 0.05\,T, which is 215\,nm.

In Fig.\,\ref{vortexLa}c we show the normalized conductance curves outside and inside the vortex core. Outside the vortex core, we observe the usual superconducting gap. At the centre of the vortex core, the superconducting gap vanishes and we observe instead a broad peak in the tunneling conductance centred at zero bias.

Due to multiple Andreev scattering and the singular shape of the order parameter at the vortex core, quantized states can form inside vortices, particularly, when the ratio $\Delta^2/$E$_F$ is large. These states are called Caroli de Gennes Matricon (CdGM) states\,\cite{Osti_4055167}. LaRu$_2$P$_2$ does not present correlations, for which E$_F$ is relatively large (0.58\,eV). As $\Delta$ is relatively small (0.61\,meV), $\Delta^2/$E$_F$ is small in LaRu$_2$P$_2$. In such a situation, CdGM vortex cores states are averaged into a single sharp peak at the vortex core, as observed for instance in 2H-NbSe$_2$\,\cite{Physrevlett.64.2711,Physrevlett.62.214,Hayashi1998,Guillamon2008superconducting}. Contrary to 2H-NbSe$_2$ and related systems, in LaRu$_2$P$_2$, the electronic mean free path $\ell \le \xi$, so that the peak is strongly broadened by impurity scattering\,\cite{Guillamon2008superconducting,Physrevb.97.134501,Hayashi1998}.

To quantify the size of the vortex core in LaRu$_2$P$_2$, we followed the method described in Ref.\,\cite{Physrevb.71.134505,Physrevb.94.014517}. In Fig.\,\ref{vortexLa}d we show the radially averaged and normalized tunneling conductance around two vortices at 0.05\,T and 0.07\,T. From these profiles, we extract the vortex core size and plot it against the magnetic field in Fig.\,\ref{vortexLa}e. We normalize the magnetic field to H$_{c2}$ and compare the same quantity for other superconducting compounds\,\cite{PhysRevB.92.054507}. The vortices in LaRu$_2$P$_2$ are by more than an order of magnitude larger than the vortices in other materials. From the extrapolation of the vortex core size to H$_{c2}$ we obtain for the superconducting coherence length $\xi \simeq$\,52\,nm. This value is in good agreement with the value of $\xi$ obtained through macroscopic measurements ($\xi \approx$\,50\,nm, see Ref.\,\cite{Li2016pressure}).

Thus, our results show that LaRu$_2$P$_2$ is a s-wave BCS superconductor with a low critical field and a large coherence length. Vortices present localized CdGM states, in agreement with s-wave superconductivity. Likely, it is a phonon-mediated superconductor with a rather isotropic superconducting gap structure.

The Eliashberg spectral function has been obtained for LaRu$_2$P$_2$ in Ref.\,\cite{Karaca2016first}. The Eliashberg spectral function describes the probability distribution of phonon frequencies and the strength of their coupling with electrons. The T$_c$ calculated in Ref.\,\cite{Karaca2016first} is very close to the actual observed T$_c$. The phonon structure in LaRu$_2$P$_2$  shows two soft modes out of the a-b plane. The electron phonon coupling for these soft modes is particularly large because of their coupling to the Ru-4d orbitals, presenting a high electronic density of states\,\cite{Karaca2016first}. These modes might soften further under pressure and be related to the collapsed tetragonal transition observed above 2 GPa, which renders LaRu$_2$P$_2$ non superconducting\,\cite{ Li2016pressure,Foroozani2014hydrostatic}.

Often, well defined directional soft modes produce anisotropic gap structures and are observed in the tunneling spectroscopy\,\cite{Physrevb.64.020503}. The Ru-4d character is spread out over the Fermi surface, which eliminates anisotropies, and probably favors a homogeneous superconducting gap. As we show here, the superconducting gap of LaRu$_2$P$_2$ is rather isotropic over the Fermi surface, suggesting that electron-phonon coupling is also isotropic. All this makes it very difficult to identify phonon modes the tunneling spectroscopy.

On the other hand, LaRu$_2$P$_2$ lacks a two-dimensional hole pocket at the center of the Brillouin zone\,\cite{Razzoli2012bulk, Moll2011quantum}. The multiple values of the superconducting gap often found in Fe based materials\,\cite{Physrevb.97.134501,Doi:10.1126/science.aal1575} are related to the set of two-dimensional pockets and and the coexistence with other order such as antiferromagnetism (AFM) with wavevectors that connect the pockets at the centre and at the borders of the Brillouin zone. Our results suggest that the interactions among these two sets of bands are crucial to create conditions favouring superconductivity with largely different superconducting gap values. This suggests, in turn, that multiple gap, high critical temperature and s$^{+-}$ superconductivity in the Fe based pnictides arise from electronic interactions between different bands crossing the Fermi level.

In summary, we have determined the superconducting density of states and observed vortices in LaRu$_2$P$_2$ using STM. We find that there are no states at the Fermi level and no signatures for a large distribution of values of the superconducting gap. The gap follows the temperature dependence expected from s-wave BCS theory. From the vortex core size, we determine the coherence length ($\xi  \approx$ 50 nm). Our measurements show that the microscopic properties of LaRu$_2$P$_2$ are very similar to those of electron-phonon mediated intermetallic superconductors.

\section*{Acknowledgments} 

This work was supported by the Spanish Research State Agency (PID2020-114071RB-I00, CEX2023-001316-M, TED2021-130546B-I00), by the Comunidad de Madrid through program NANOMAGCOST-CM (Program No.S2018/NMT-4321) and by the European Research Council PNICTEYES grant agreement 679080  and by VectorFieldImaging grant agreement 101069239. We acknowledge collaborations through EU program Cost CA21144 (Superqumap). We acknowledge segainvex for the support in cryogenics and for the design and construction of STM equipment. We also thank Rafael \'Alvarez Montoya and Jos\'e Mar\'ia Castilla for technical support.



%

\end{document}